\documentclass[twocolumn,10pt]{article}
\makeindex
\usepackage{epsf}
\usepackage{axodraw}
\usepackage{epsfig}                            
\usepackage{graphicx}
\usepackage{rotate}
\usepackage{latexsym}
\renewcommand{\cal}{\mathcal}

\def\sloppy{\tolerance=100000\hfuzz=\maxdimen\vfuzz=\maxdimen}
\begin{document}

\newcommand{\spro}[2]{{#1}\cdot{#2}}
\newcommand{\ds}{\displaystyle}
\newcommand{\Ddrh}{{\ds\frac{1}{\hat{\varepsilon}}}}
\newcommand{\DiagramFermionToBosonFullWithMomenta}[8][70]{
  \vcenter{\hbox{
  \SetScale{0.8}
  \begin{picture}(#1,50)(15,15)
    \put(27,22){$\nearrow$}      
    \put(27,54){$\searrow$}    
    \put(59,29){$\to$}    
    \ArrowLine(25,25)(50,50)      \Text(34,20)[lc]{#6} \Text(11,20)[lc]{#3}
    \ArrowLine(50,50)(25,75)      \Text(34,60)[lc]{#7} \Text(11,60)[lc]{#4}
    \Photon(50,50)(90,50){2}{8}   \Text(80,40)[lc]{#2} \Text(55,33)[ct]{#8}
    \Vertex(50,50){2,5}          \Text(60,48)[cb]{#5} 
    \Vertex(90,50){2}
  \end{picture}}}
  }
\newcommand{\DiagramFermionToBosonPropagator}[4][85]{
  \vcenter{\hbox{
  \SetScale{0.8}
  \begin{picture}(#1,50)(15,15)
    \ArrowLine(25,25)(50,50)
    \ArrowLine(50,50)(25,75)
    \Photon(50,50)(105,50){2}{8}   \Text(90,40)[lc]{#2}
    \Vertex(50,50){0.5}         \Text(80,48)[cb]{#3}
    \GCirc(82,50){8}{1}            \Text(55,48)[cb]{#4}
    \Vertex(105,50){2}
  \end{picture}}}
  }
\newcommand{\DiagramFermionToBosonEffective}[3][70]{
  \vcenter{\hbox{
  \SetScale{0.8}
  \begin{picture}(#1,50)(15,15)
    \ArrowLine(25,25)(50,50)
    \ArrowLine(50,50)(25,75)
    \Photon(50,50)(90,50){2}{8}   \Text(80,40)[lc]{#2}
    \BBoxc(50,50)(5,5)            \Text(55,48)[cb]{#3}
    \Vertex(90,50){2}
  \end{picture}}}
  }
\newcommand{\DiagramFermionToBosonFull}[3][70]{
  \vcenter{\hbox{
  \SetScale{0.8}
  \begin{picture}(#1,50)(15,15)
    \ArrowLine(25,25)(50,50)
    \ArrowLine(50,50)(25,75)
    \Photon(50,50)(90,50){2}{8}   \Text(80,40)[lc]{#2}
    \Vertex(50,50){2.5}          \Text(60,48)[cb]{#3}
    \Vertex(90,50){2}
  \end{picture}}}
  }
\newcommand{\pc}{\,\%}
\newcommand{\bffi}[1]{{\overline{f}}_{#1}}
\newcommand{\ffi}[1]{f_{#1}}
\newcommand{\Cclu}{{\Black{$\clubsuit$}}}
\newcommand{\Cdia}{{\Red{$\diamondsuit$}}}
\newcommand{\Chea}{{\Red{$\heartsuit$}}}
\newcommand{\Cspa}{{\Black{$\spadesuit$}}}
\newcommand{\Cosl}{{\Red{$\oslash$}}}
\newcommand{\Cnab}{{\Red{$\nabla$}}}
\newcommand{\Crhd}{{\Red{$\rhd$}}}
\newcommand{\Csha}{{\Red{$\sharp$}}}
\newcommand{\Cbtr}{{\Red{$\bigtriangleup$}}}
\newcommand{\Cbox}{{\Red{$\Box$}}}
\newcommand{\Codo}{{\Red{$\odot$}}}
\newcommand{\Cfla}{{\Red{$\flat$}}}
\newcommand{\Cnat}{{\Red{$\natural$}}}
\newcommand{\Cfro}{{\Red{$\frown$}}}
\newcommand{\Csmi}{{\Red{$\smile$}}}
\newcommand{\Comi}{{\Red{$\ominus$}}}
\newcommand{\bec}{\begin{center}}
\newcommand{\eec}{\end{center}}
\newcommand{\nBL}{\Black}
\newcommand{\nBl}{\Blue}
\newcommand{\nRe}{\Red}
\newcommand{\nBr}{\Brown}
\newcommand{\nOr}{\Orange}
\newcommand{\nCy}{\Cyan}
\newcommand{\nMa}{\Magenta}
\newcommand{\nGr}{\Green}
\newcommand{\bBL}{\bf\Black}
\newcommand{\bBl}{\bf\Blue}
\newcommand{\bRe}{\bf\Red}
\newcommand{\bBr}{\bf\Brown}
\newcommand{\bOr}{\bf\Orange}
\newcommand{\bCy}{\bf\Cyan}
\newcommand{\bMa}{\bf\Magenta}
\newcommand{\bGr}{\bf\Green}
\newcommand{\eBL}{\em\Black}
\newcommand{\eBl}{\em\Blue}
\newcommand{\eRe}{\em\Red}
\newcommand{\eBr}{\em\Brown}
\newcommand{\eOr}{\em\Orange}
\newcommand{\eCy}{\em\Cyan}
\newcommand{\eMa}{\em\Magenta}
\newcommand{\eGr}{\em\Green}
\newcommand{\vj}[4]{{\sl #1~}{\bf #2 }\ifnum#3<100 (19#3) \else (#3) \fi #4}
\newcommand{\ej}[3]{{\bf #1~}\ifnum#2<100 (19#2) \else (#2) \fi #3}
\newcommand{\ap}[3]{\vj{Ann.~Phys.}{#1}{#2}{#3}}
\newcommand{\app}[3]{\vj{Acta~Phys.~Pol.}{#1}{#2}{#3}}
\newcommand{\cmp}[3]{\vj{Commun. Math. Phys.}{#1}{#2}{#3}}
\newcommand{\cnpp}[3]{\vj{Comments Nucl. Part. Phys.}{#1}{#2}{#3}}
\newcommand{\cpc}[3]{\vj{Comp. Phys. Commun.}{#1}{#2}{#3}}
\newcommand{\epj}[3]{\vj{Eur. Phys. J.}{#1}{#2}{#3}}
\newcommand{\hpa}[3]{\vj{Helv. Phys.~Acta}{#1}{#2}{#3}} 
\newcommand{\ijmp}[3]{\vj{Int. J. Mod. Phys.}{#1}{#2}{#3}}
\newcommand{\jetp}[3]{\vj{JETP}{#1}{#2}{#3}}
\newcommand{\jetpl}[3]{\vj{JETP Lett.}{#1}{#2}{#3}}
\newcommand{\jmp}[3]{\vj{J. Math. Phys.}{#1}{#2}{#3}}
\newcommand{\jp}[3]{\vj{J. Phys.}{#1}{#2}{#3}}
\newcommand{\lnc}[3]{\vj{Lett. Nuovo Cimento}{#1}{#2}{#3}}
\newcommand{\mpl}[3]{\vj{Mod. Phys. Lett.}{#1}{#2}{#3}}
\newcommand{\nc}[3]{\vj{Nuovo Cimento}{#1}{#2}{#3}}
\newcommand{\nim}[3]{\vj{Nucl. Instr. Meth.}{#1}{#2}{#3}}
\newcommand{\np}[3]{\vj{Nucl. Phys.}{#1}{#2}{#3}}
\newcommand{\pl}[3]{\vj{Phys. Lett.}{#1}{#2}{#3}}
\newcommand{\prp}[3]{\vj{Phys. Rep.}{#1}{#2}{#3}}
\newcommand{\pr}[3]{\vj{Phys. Rev.}{#1}{#2}{#3}}
\newcommand{\prl}[3]{\vj{Phys. Rev. Lett.}{#1}{#2}{#3}}
\newcommand{\ptp}[3]{\vj{Prog. Theor. Phys.}{#1}{#2}{#3}}
\newcommand{\rpp}[3]{\vj{Rep. Prog. Phys.}{#1}{#2}{#3}}
\newcommand{\rmp}[3]{\vj{Rev. Mod. Phys.}{#1}{#2}{#3}}
\newcommand{\rnc}[3]{\vj{Rivista del Nuovo Cim.}{#1}{#2}{#3}}
\newcommand{\sjnp}[3]{\vj{Sov. J. Nucl. Phys.}{#1}{#2}{#3}}
\newcommand{\sptp}[3]{\vj{Suppl. Prog. Theor. Phys.}{#1}{#2}{#3}}
\newcommand{\zp}[3]{\vj{Z. Phys.}{#1}{#2}{#3}}
\newcommand{\jop}[3]{\vj{Journal of Physics} {\bf #1} (#2) #3}
\newcommand{\ibid}[3]{\vj{ibid.} {\bf #1} (#2) #3}
\newcommand{\hep}[1]{{\sl hep--ph/}{#1}}
%
%
\newcommand{\rmf}{{\rm f}}
\newcommand{\vb}{V}
\newcommand{\hq}{{\hat Q}}
\newcommand{\ben}{\begin{enumerate}}
\newcommand{\een}{\end{enumerate}}
\newcommand{\asums}[1]{\sum_{#1}}
\newcommand{\bei}{\begin{itemize}}
\newcommand{\eei}{\end{itemize}}
\newcommand{\dr}{\Delta r}
\newcommand{\ft}{t}
\newcommand{\tabn}[1]{Tab.(\ref{#1})}
\newcommand{\tabns}[2]{Tabs.(\ref{#1}--\ref{#2})}
\newcommand{\fnul}{\nu_l}
\newcommand{\fnue}{\nu_e}
\newcommand{\fnum}{\nu_{\mu}}
\newcommand{\fnut}{\nu_{\tau}}
\newcommand{\gb}{g} 
\newcommand{\ib}{i}
\newcommand{\fe}{e}
\newcommand{\ff}{f}
\newcommand{\fep}{e^{+}}
\newcommand{\fem}{e^{-}}
\newcommand{\fepm}{e^{\pm}}
\newcommand{\fp}{f^{+}}
\newcommand{\fm}{f^{-}}
\newcommand{\bqas}{\begin{eqnarray*}}
\newcommand{\eqas}{\end{eqnarray*}}
\newcommand{\scff}[1]{C_{#1}}                    
\newcommand{\TeV}{\;\mathrm{TeV}}                     
\newcommand{\gz}{\Gamma_{_{\zb}}}
\newcommand{\gw}{\Gamma_{_{\wb}}}
\newcommand{\sW}{p_{_W}}
\newcommand{\sZ}{p_{_Z}}
\newcommand{\ssp}{s_p}
\newcommand{\fW}{f_{_W}}
\newcommand{\fZ}{f_{_Z}}
\newcommand{\bzm}{M_{_0}}
\newcommand{\stw}{s_{\theta}}             
\newcommand{\ctw}{c_{\theta}}
\newcommand{\stws}{s_{\theta}^2}
\newcommand{\stwc}{s_{\theta}^3}
\newcommand{\stwf}{s_{\theta}^4}
\newcommand{\stwx}{s_{\theta}^6}
\newcommand{\ctws}{c_{\theta}^2}
\newcommand{\ctwc}{c_{\theta}^3}
\newcommand{\ctwf}{c_{\theta}^4}
\newcommand{\ctwx}{c_{\theta}^6}
\newcommand{\stwfiv}{s_{\theta}^5}
\newcommand{\ctwfiv}{c_{\theta}^5}
\newcommand{\stwsix}{s_{\theta}^6}
\newcommand{\ctwsix}{c_{\theta}^6}
\newcommand{\drii}[2]{\delta_{#1#2}}                    
\newcommand{\mlones}{m^2}
\newcommand{\Smat}{{\cal{S}}}
\newcommand{\me}{m_e}
\newcommand{\mm}{m_{\mu}}
\newcommand{\mtau}{m_{\tau}}
\newcommand{\muq}{m_u}
\newcommand{\md}{m_d}
\newcommand{\muqp}{m'_u}
\newcommand{\mdqp}{m'_d}
\newcommand{\mc}{m_c}
\newcommand{\ms}{m_s}
\newcommand{\mb}{m_b}
\newcommand{\eqn}[1]{Eq.(\ref{#1})}
\newcommand{\eqns}[2]{Eqs.(\ref{#1}--\ref{#2})}
\newcommand{\eqnss}[1]{Eqs.(\ref{#1})}
\newcommand{\eqnsc}[2]{Eqs.(\ref{#1},~\ref{#2})}
\newcommand{\tbn}[1]{Tab.~\ref{#1}}
\newcommand{\tbns}[2]{Tabs.~\ref{#1}--\ref{#2}}
\newcommand{\tbnsc}[2]{Tabs.~\ref{#1},~\ref{#2}}
\newcommand{\fig}[1]{Fig.~\ref{#1}}
\newcommand{\figs}[2]{Figs.~\ref{#1}--\ref{#2}}
\newcommand{\figsc}[2]{Figs.~\ref{#1},~\ref{#2}}
\newcommand{\sect}[1]{Sect.~\ref{#1}}
\newcommand{\subsect}[1]{Sub-Sect.~\ref{#1}}
\newcommand{\subsects}[2]{Sub-Sects.~\ref{#1},~\ref{#2}}
\newcommand{\pmom}{p}
\newcommand{\pmomp}{p'}
\newcommand{\pmoms}{p^2}
\newcommand{\pmomq}{p^4}
\newcommand{\pmomx}{p^6}
\newcommand{\pmomi}[1]{p_{#1}}
\newcommand{\pmomis}[1]{p^2_{#1}}
\newcommand{\Glone}{\Gamma}
\newcommand{\gbs}{g^2}
\newcommand{\gbc}{g^3}
\newcommand{\gbf}{g^4}
\newcommand{\sgh}{{\hat\Sigma}}
\newcommand{\Pgg}{\Pi_{\ph\ph}}
\newcommand{\Ptg}{\Pi_{_{3Q}}}
\newcommand{\Ptt}{\Pi_{_{33}}}
\newcommand{\Pzg}{\Pi_{_{\zb\ab}}}
\newcommand{\Pzga}[2]{\Pi^{#1}_{_{\zb\ab}}\lpar#2\rpar}
\newcommand{\Pf}{\Pi_{_F}}
\newcommand{\Sgg}{\Sigma_{_{\ab\ab}}}
\newcommand{\Szg}{\Sigma_{_{\zb\ab}}}
\newcommand{\SVV}{\Sigma_{_{\vb\vb}}}
\newcommand{\USvv}{{\hat\Sigma}_{_{\vb\vb}}}
\newcommand{\Sww}{\Sigma_{_{\wb\wb}}}
\newcommand{\Swwg}{\Sigma^{_G}_{_{\wb\wb}}}
\newcommand{\Szz}{\Sigma_{_{\zb\zb}}}
\newcommand{\Shh}{\Sigma_{_{\hb\hb}}}
\newcommand{\Spzz}{\Sigma'_{_{\zb\zb}}}
\newcommand{\Stg}{\Sigma_{_{3Q}}}
\newcommand{\Stt}{\Sigma_{_{33}}}
\newcommand{\bSww}{{\overline\Sigma}_{_{WW}}}
\newcommand{\bStg}{{\overline\Sigma}_{_{3Q}}}
\newcommand{\bStt}{{\overline\Sigma}_{_{33}}}
\newcommand{\sman}{s}
\newcommand{\smanp}{s'}
\newcommand{\tman}{t}
\newcommand{\uman}{u}
\newcommand{\smani}[1]{s_{#1}}
\newcommand{\bsmani}[1]{{\bar{s}}_{#1}}
\newcommand{\smans}{s^2}
\newcommand{\tmans}{t^2}
\newcommand{\umans}{u^2}
\newcommand{\ec}{e}
\newcommand{\ecs}{e^2}
\newcommand{\ect}{e^3}
\newcommand{\ecq}{e^4}
\newcommand{\Reb}{{\rm{Re}}}
\newcommand{\Imb}{{\rm{Im}}}
\newcommand{\sany}{s}
\newcommand{\cany}{c}
\newcommand{\sanys}{s^2}
\newcommand{\canys}{c^2}
\newcommand{\scats}{s}
\newcommand{\scatss}{s^2}
\newcommand{\scatsi}[1]{s_{#1}}
\newcommand{\scatsis}[1]{s^2_{#1}}
\newcommand{\scatst}[2]{s_{#1}^{#2}}
\newcommand{\scatc}{c}
\newcommand{\scatcs}{c^2}
\newcommand{\scatci}[1]{c_{#1}}
\newcommand{\scatcis}[1]{c^2_{#1}}
\newcommand{\scatct}[2]{c_{#1}^{#2}}
\newcommand{\bq}{\begin{equation}}                   
\newcommand{\eq}{\end{equation}}
\newcommand{\bqa}{\begin{eqnarray}}
\newcommand{\eqa}{\end{eqnarray}}
\newcommand{\ba}[1]{\begin{array}{#1}}
\newcommand{\ea}{\end{array}}
\newcommand{\lpar}{\left(}                            
\newcommand{\rpar}{\right)} 
\newcommand{\lrbr}{\left[}
\newcommand{\rrbr}{\right]}
\newcommand{\lcbr}{\left\{}
\newcommand{\rcbr}{\right\}} 
\newcommand{\ph}{\gamma}
\newcommand{\zb}{Z}
\newcommand{\wb}{W}            
\newcommand{\wbp}{W^{+}}
\newcommand{\wbm}{W^{-}}
\newcommand{\wbpm}{W^{\pm}}
\newcommand{\barf}{\overline f}                
\newcommand{\barl}{\overline l}
\newcommand{\barq}{\overline q}
\newcommand{\barqp}{\overline{q}'}
\newcommand{\barb}{\overline b}
\newcommand{\bart}{\overline t}
\newcommand{\barc}{\overline c}
\newcommand{\baru}{\overline u}
\newcommand{\bard}{\overline d}
\newcommand{\bars}{\overline s}
\newcommand{\barv}{\overline v}
\newcommand{\barnu}{\overline{\nu}}
\newcommand{\barne}{\overline{\nu}_{\fe}}
\newcommand{\barnm}{\overline{\nu}_{\flm}}
\newcommand{\barnt}{\overline{\nu}_{\flt}}
\newcommand{\mh}{M_{_H}}
\newcommand{\mls}{m^2_l}
\newcommand{\mVs}{M^2_{_V}}
\newcommand{\mw}{M_{_W}}
\newcommand{\mws}{M^2_{_W}}
\newcommand{\mwc}{M^3_{_W}}
\newcommand{\LMs}{M^2}
\newcommand{\LMc}{M^3}
\newcommand{\mz}{M_{_Z}}
\newcommand{\mzs}{M^2_{_Z}}
\newcommand{\mzc}{M^3_{_Z}}
\newcommand{\bzms}{M^2_{_0}}
\newcommand{\bzmc}{M^3_{_0}}
\newcommand{\bhms}{M^2_{_{0H}}}
\newcommand{\mhs}{M^2_{_H}}
\newcommand{\mfs}{m^2_f}
\newcommand{\mfc}{m^3_f}
\newcommand{\mfps}{m^2_{f'}}
\newcommand{\mfhs}{m^2_{h}}
\newcommand{\mfpc}{m^3_{f'}}
\newcommand{\mts}{m^2_t}
\newcommand{\mes}{m^2_e}
\newcommand{\mms}{m^2_{\mu}}
\newcommand{\mmc}{m^3_{\mu}}
\newcommand{\mmfour}{m^4_{\mu}}
\newcommand{\mmf}{m^5_{\mu}}
\newcommand{\mmfive}{m^5_{\mu}}
\newcommand{\mmsix}{m^6_{\mu}}
\newcommand{\mminv}{\frac{1}{m_{\mu}}}
\newcommand{\mtaus}{m^2_{\tau}}
\newcommand{\mus}{m^2_u}
\newcommand{\mds}{m^2_d}
\newcommand{\muqps}{m'^2_u}
\newcommand{\mdqps}{m'^2_d}
\newcommand{\mcs}{m^2_c}
\newcommand{\mss}{m^2_s}
\newcommand{\mbs}{m^2_b}
\newcommand{\mups}{M^2_u}
\newcommand{\mdps}{M^2_d}
\newcommand{\mcps}{M^2_c}
\newcommand{\msps}{M^2_s}
\newcommand{\mbps}{M^2_b}
\newcommand{\mt}{m_t}
\newcommand{\gf}{G_{\ssF}}
\newcommand{\ssZ}{{\scriptscriptstyle{\zb}}}
\newcommand{\ssW}{{\scriptscriptstyle{\wb}}}
\newcommand{\ssH}{{\scriptscriptstyle{\hb}}}
\newcommand{\ssV}{{\scriptscriptstyle{\vb}}}
\newcommand{\ssA}{{\scriptscriptstyle{A}}}
\newcommand{\ssB}{{\scriptscriptstyle{B}}}
\newcommand{\ssC}{{\scriptscriptstyle{C}}}
\newcommand{\ssD}{{\scriptscriptstyle{D}}}
\newcommand{\ssF}{{\scriptscriptstyle{F}}}
\newcommand{\ssG}{{\scriptscriptstyle{G}}}
\newcommand{\ssL}{{\scriptscriptstyle{L}}}
\newcommand{\ssM}{{\scriptscriptstyle{M}}}
\newcommand{\ssN}{{\scriptscriptstyle{N}}}
\newcommand{\ssP}{{\scriptscriptstyle{P}}}
\newcommand{\ssQ}{{\scriptscriptstyle{Q}}}
\newcommand{\ssR}{{\scriptscriptstyle{R}}}
\newcommand{\ssS}{{\scriptscriptstyle{S}}}
\newcommand{\ssT}{{\scriptscriptstyle{T}}}
\newcommand{\ssU}{{\scriptscriptstyle{U}}}
\newcommand{\ssX}{{\scriptscriptstyle{X}}}
\newcommand{\ssY}{{\scriptscriptstyle{Y}}}
\newcommand{\ssI}{{\scriptscriptstyle{I}}}
\newcommand{\ord}[1]{{\cal O}\lpar#1\rpar}
\newcommand{\als}{\alpha_{_S}}
%
%
\def\beq{\begin{equation}}
\def\eeq{\end{equation}}
\def\beqar{\begin{eqnarray}}
\def\eeqar{\end{eqnarray}}
\def\barr#1{\begin{array}{#1}}
\def\earr{\end{array}}
\def\bfi{\begin{figure}}
\def\efi{\end{figure}}
\def\btab{\begin{table}}
\def\etab{\end{table}}
\def\bce{\begin{center}}
\def\ece{\end{center}}
\def\nn{\nonumber}
\def\nl{\nonumber\\}

\def\al{\alpha}
\def\be{\beta}
\def\ga{\gamma}

\def\refeq#1{\mbox{(\ref{#1})}}
\def\reffi#1{\mbox{Figure~\ref{#1}}}
\def\reffis#1{\mbox{Figures~\ref{#1}}}
\def\refta#1{\mbox{Table~\ref{#1}}}
\def\reftas#1{\mbox{Tables~\ref{#1}}}
\def\citere#1{\mbox{Ref.~\cite{#1}}}
\def\citeres#1{\mbox{Refs.~\cite{#1}}}

\newcommand{\GeV}{\unskip\,\mathrm{GeV}}
\newcommand{\MeV}{\unskip\,\mathrm{MeV}}

\newcommand{\ri}{{\mathrm{i}}}

\renewcommand{\O}{{\cal O}}
\newcommand{\M}{{\cal{M}}}

\def\mathswitchr#1{\relax\ifmmode{\mathrm{#1}}\else$\mathrm{#1}$\fi}
\newcommand{\PW}{\mathswitchr W}
\newcommand{\PZ}{\mathswitchr Z}
\newcommand{\PH}{\mathswitchr H}
\newcommand{\Pe}{\mathswitchr e}
\newcommand{\Pd}{\mathswitchr d}
\newcommand{\Pu}{\mathswitchr u}
\newcommand{\Ps}{\mathswitchr s}
\newcommand{\Pc}{\mathswitchr c}
\newcommand{\Pb}{\mathswitchr b}
\newcommand{\Pt}{\mathswitchr t}
\newcommand{\Pep}{\mathswitchr {e^+}}
\newcommand{\Pem}{\mathswitchr {e^-}}
\newcommand{\PWp}{\mathswitchr {W^+}}
\newcommand{\PWm}{\mathswitchr {W^-}}

\def\mathswitch#1{\relax\ifmmode#1\else$#1$\fi}
\newcommand{\MW}{\mathswitch {M_\PW}}
\newcommand{\MZ}{\mathswitch {M_\PZ}}
\newcommand{\MH}{\mathswitch {M_\PH}}
\newcommand{\Me}{\mathswitch {m_\Pe}}
\newcommand{\GW}{\Gamma_{\PW}}
\newcommand{\GZ}{\Gamma_{\PZ}}

\newcommand{\cw}{\mathswitch {c_\mathrm{w}}}
\newcommand{\sw}{\mathswitch {s_\mathrm{w}}}
\newcommand{\GF}{\mathswitch {G_\mu}}

\newcommand{\lsim}
{\mathrel{\raisebox{-.3em}{$\stackrel{\displaystyle <}{\sim}$}}}
\newcommand{\gsim}
{\mathrel{\raisebox{-.3em}{$\stackrel{\displaystyle >}{\sim}$}}}

\def\ie{i.e.\ }
\def\eg{e.g.\ }
\def\cf{cf.\ }

\newcommand{\DPA}{{\mathrm{DPA}}}
\newcommand{\virt}{{\mathrm{virt}}}
\newcommand{\fact}{{\mathrm{fact}}}
\newcommand{\nonfact}{{\mathrm{nfact}}}

\newcommand{\eeWW}{{\Pe^+ \Pe^-\to \PW^+ \PW^-}}
\newcommand{\Wpff}{{\PW^+ \to f_1\bar f_2}}
\newcommand{\Wmff}{{\PW^- \to f_3\bar f_4}}
\newcommand{\eeWWffff}{\Pep\Pem\to\PW\PW\to 4f}
\newcommand{\eeffff}{\Pep\Pem\to 4f}
\newcommand{\eeffffg}{\eeffff\ga}

\newcommand{\mpar}[1]{{\marginpar{\hbadness10000%
                      \sloppy\hfuzz10pt\boldmath\bf#1}}%
                      \typeout{marginpar: #1}\ignorespaces}
\def\draftdate{\relax}
\def\mua{\marginpar[\boldmath\hfil$\uparrow$]%
                   {\boldmath$\uparrow$\hfil}%
                    \typeout{marginpar: $\uparrow$}\ignorespaces}
\def\mda{\marginpar[\boldmath\hfil$\downarrow$]%
                   {\boldmath$\downarrow$\hfil}%
                    \typeout{marginpar: $\downarrow$}\ignorespaces}
\def\mla{\marginpar[\boldmath\hfil$\rightarrow$]%
                   {\boldmath$\leftarrow $\hfil}%
                    \typeout{marginpar: $\leftrightarrow$}\ignorespaces}
\def\Mua{\marginpar[\boldmath\hfil$\Uparrow$]%
                   {\boldmath$\Uparrow$\hfil}%
                    \typeout{marginpar: $\Uparrow$}\ignorespaces}
\def\Mda{\marginpar[\boldmath\hfil$\Downarrow$]%
                   {\boldmath$\Downarrow$\hfil}%
                    \typeout{marginpar: $\Downarrow$}\ignorespaces}
\def\Mla{\marginpar[\boldmath\hfil$\Rightarrow$]%
                   {\boldmath$\Leftarrow $\hfil}%
                    \typeout{marginpar: $\Leftrightarrow$}\ignorespaces}
%
%
\newcommand{\sss}[1]{\mbox{\scriptsize #1}}
\newcommand{\real}{{\rm{Re}}}
\newcommand{\imag}{{\rm{Im}}}
\newcommand{\J}{{\cal J}}
\newcommand{\I}{{\cal I}}
\newcommand{\OO}{{\cal O}}
\newcommand{\PP}{{\cal P}}
\newcommand{\D}{{\cal D}}
\newcommand{\T}{\mbox{T}}
\newcommand{\Tr}{\mbox{Tr}\,}
\newcommand{\F}{\mbox{F}}
\newcommand{\BBCG}{\mbox{G}}
\newcommand{\TF}{\mbox{\boldmath $F$}}
\newcommand{\TA}{\mbox{\boldmath $A$}}
\newcommand{\LL}{{\cal L}}
\newcommand{\LNL}{{\cal L}_{\sss{NL}}}
\newcommand{\SNL}{{\cal S}_{\sss{NL}}}
\newcommand{\SL}{{\cal S}_{\sss{L}}}
\newcommand{\Ds}{D\hspace{-0.63em}/\hspace{0.1em}}
\newcommand{\TAs}{\TA\hspace{-0.65em}/\hspace{0.1em}}
\newcommand{\partials}{\partial\hspace{-0.50em}/\hspace{0.1em}}
%
%
\newcommand{\bfig}{\begin{center}\begin{picture}}
\newcommand{\efig}[1]{\end{picture}\\{\small #1}\end{center}}
\newcommand{\flin}[2]{\ArrowLine(#1)(#2)}
\newcommand{\wlin}[2]{\DashLine(#1)(#2){2.5}}
\newcommand{\zlin}[2]{\DashLine(#1)(#2){5}}
\newcommand{\glin}[3]{\Photon(#1)(#2){2}{#3}}
\newcommand{\lin}[2]{\Line(#1)(#2)}
\newcommand{\sof}{\SetOffset}
\newcommand{\bmip}[2]{\begin{minipage}[t]{#1pt}\bfig(#1,#2)}
\newcommand{\emip}[1]{\efig{#1}\end{minipage}}
\newcommand{\putk}[2]{\Text(#1)[r]{$p_{#2}$}}
\newcommand{\putp}[2]{\Text(#1)[l]{$p_{#2}$}}
\newcommand{\ibidem}{{\it ibidem\/},}
\newcommand{\vpb}{}
\newcommand{\p}[1]{{\scriptstyle{\,(#1)}}}
%
%
\newcommand{\parent}[1]{\lpar#1\rpar}
\newcommand{\rbrak}[1]{\lrbr#1\rrbr}
\newcommand{\ra}{\rightarrow}
\newcommand{\beanon}{\begin{eqnarray*}}
\newcommand{\eeanon}{\end{eqnarray*}}
\newcommand{\ul}{\underline}
\newcommand{\ol}{\overline}
\newcommand{\dotp}{\!\cdot\!}
\newcommand{\thet}[1]{\theta\lpar#1\rpar}
\newcommand{\delt}[1]{\delta\lpar#1\rpar}
\newcommand{\gtap}{\stackrel{\displaystyle >}{\,_{\! \,_{\displaystyle
\sim}}}}  
\newcommand{\ltap}{\stackrel{\displaystyle <}{\,_{\! \,_{\displaystyle
\sim}}}}  
\newcommand{\umu}{^{\mu}}
\newcommand{\lmu}{_{\mu}}
\newcommand{\ub}{\bar{u}}
\newcommand{\vbar}{\bar{v}}
\newcommand{\gp}{(1+\gamma^5)}
\newcommand{\ep}{\epsilon}
\newcommand{\sla}[1]{/\!\!\!#1}
\newcommand{\suml}{\sum\limits}
\renewcommand{\to}{\rightarrow}
\newcommand{\unity}{1\!\!1}
\renewcommand{\iff}{\;\;\Longleftrightarrow\;\;}
\newcommand{\gsovermu}{\kappa}
\newcommand{\mut}{m_t^2}
\newcommand{\gmw}{\Gamma_{_W}}
\newcommand{\gmz}{\Gamma_{_Z}}
\newcommand\pb{\;[\mathrm{pb}]}
\newcommand{\processccten}{$e^-e^+\to \mu^-\bar{\nu}_\mu u\bar{d}$}
\newcommand{\processcceleven}{$e^-e^+\to s\bar{c} u\bar{d}$}
\newcommand{\processcctwenty}{$e^-e^+\to e^-\bar{\nu}_eu\bar{d}$}
\newcommand{\processccemu}{$e^-e^+\to e^-\bar{\nu}_e\mu^+\nu_\mu$}
\newcommand{\processmixeevv}{$e^-e^+\to e^-e^+\nu_e\bar{\nu}_e$}
\newcommand{\processnceevv}{$e^-e^+\to e^-e^+\nu_\mu\bar{\nu}_\mu$}
\newcommand{\wph}{{\tt WPHACT}}
\newcommand{\wto}{{\tt WTO}}
\newcommand{\com}[1] {{(\bf #1})}
%
%
\def\Was{W\c as}
\def\Order#1{${\cal O}(#1$)}
\def\Ordpr#1{${\cal O}(#1)_{prag}$}
\def\bbe{\bar{\beta}}
\def\tbe{\tilde{\beta}}
\def\tal{\tilde{\alpha}}
\def\tom{\tilde{\omega}}
\def\half{ {1\over 2} }
\def\alf1{ {\alpha\over\pi} }

\title{Recent Theoretical Developments in LEP~2 Physics}

\author{Giampiero Passarino \\[5mm]
Dipartimento di Fisica Teorica, Universit\`a di Torino, Italy\\
INFN, Sezione di Torino, Italy\\
E-mail: giampiero@to.infn.it}

\twocolumn[\maketitle\abstract{
Recent theoretical developments in $e^+e^-$-annihilation into fermion
pairs are summarized. In particular, two-fermion production,
DPA for $\wb-\wb$ signal, single-$\wb$ production and $\zb-\zb$ signal
\\[1cm]}]

After an illustrious career LEP stops running rather soon so it is unlikely 
there will be any more 
data in this energy region and we all must try to do the best we can to get 
the most accurate measurements and the most precise predictions we can.
From the point of view of theory there is of course no deep 
reason why the theory uncertainty should be reduced below that of the 
experimental precision, but it is surely a useful target as the theory error
has to be added in quadrature in looking for deviations from the standard 
model. 

In this talk the most recent theoretical developments connected with LEP~2
physics will be shortly reviewed. As the LEP~2 community has written
a report that has just come out I refer the interested reader to that
report~\cite{YR} where one of the goals was to summarize and review critically 
the progress made in theoretical calculations and their implementation in 
computer programs since the 1995 workshop on {\it Physics at LEP2}.

\begin{itemize}
\item $e^+e^- \to \barf f (\ph, {\rm pairs})$
\end{itemize}

On the basis of comparisons of various calculations, theoretical
uncertainties have been estimated and compared with those for the final
LEP~2 data analysis.
In the following list we summarize the present status of theoretical 
and experimental accuracy as given in the report of the 2f Working Group of 
the LEP~2/MC Workshop~\cite{2f} to which we refer for more details:

\begin{enumerate}

\item $e+e^- \to \barq q (\ph)$ 
  { 0.3\,\%} /  { 0.1\,\%-0.2\,\%} 
\item $e+e^- \to \mu^+\mu^- (\ph)$      
  { 0.4\,\%} /  { 0.4\,\%-0.5\,\%} 
\item  $e+e^- \to \tau^+\tau^- (\ph)$    
  { 0.4\,\%} /  { 0.4\,\%-0.6\,\%} 
\item  $e+e^- \to e^+e^- (\ph)$ (endcap) 
  { 0.5\,\%} /  { 0.1\,\%}  
\item $e+e^- \to e^+e^- (\ph)$ (barrel) 
  { 2.0\,\%} /  { 0.2\,\%}    
\item  $e+e^- \to e^+e^- (\ph)$          
  { 3.0\,\%} /  { 1.5\,\%}  
\item $e+e^- \to {\rm l}^+{\rm l}^-$    
  { 1.0\,\%} /  { 0.5\,\%}  
\item  $e+e^- \to \barnu \nu (\ph)$      
  { 4.0\,\%} /  { 0.5\,\%}  

\end{enumerate}

First entry is the present theoretical uncertainty, second one is the 
experimental precision tag.
The total hadronic and leptonic cross-sections are now
predicted to the total precision tag of $0.2\,\%$,
(excluding pairs) by ZFITTER~\cite{zfitter} and KKMC~\cite{kkmc}. 

\begin{itemize}
\item News for Pairs in $e^+e^-$ annihilation
\end{itemize}

Shortly before and during this workshop a lot of new codes for 
pair corrections at LEP~2 were developed.
Before 1999, only the diagram-based pair correction with 
$\smanp = M_{\rm prop}^2$ could be calculated by ZFITTER and 
TOPAZ0~\cite{topaz0}.

Common exponentiation of IS-$\ph$ and ISNS$_{\gamma}$ pairs for energies 
away from the $\zb$-peak as well as optional ISS$_{\gamma}$ pairs were 
implemented in both codes in 1999 (see~\cite{2f} for their definition).
Now ZFITTER has been upgraded to include explicit 
FS$_\gamma$ with the possibility of mass cuts.  
Furthermore, the new GENTLE/4fan~\cite{gentle} offers even
more options with mass cuts on all pairs and inclusion of pairs from virtual 
$\zb$ and swapped FS diagrams and a new combination of KKMC and KORALW is 
being developed.

The main achievements in this area can be summarized as follows:
a proposal for a signal definition which can be, to 
better than $0.1\%$ accuracy defined either based on cuts or on diagrams.
The determination of efficiency corrections using full 
event generators has been checked for GRC4f~\cite{grace} to a precision of 
$0.1\%$, from a comparison of real pair cross-sections with 
GENTLE.
However, problems of pairing ambiguities for four identical fermions 
become increasingly important with the larger $\zb\zb$ cross-sections 
at high energies.
From varying pairing algorithms, a worst-case difference of $0.8$ 
per mill was found for inclusive hadrons at $206\,$GeV. 
Furthermore, differences for pair corrections
between $\smanp$ definitions via the propagator or primary pair mass
in the diagram-based approach have been determined and
GENTLE -- ZFITTER both find them to be about $0.3 (1.1)$ per mill for 
high $\smanp$ hadrons (muons). 

Maximum differences for the  diagram-based pair correction of 
$1.7 (1.5)$ per mill for inclusive hadrons (muons) and
$0.2 (0.4)$ per mill for  high $\smanp$ hadrons (muons) 
between any two of 
the programs GENTLE, ZFITTER and TOPAZ0 have been found.
Compared to the LEP-combined statistical precision of the
measurements all these differences are small. Even the $1.7$ per mill
difference is only about half of the expected LEP-combined statistical error.

Finally, a first complete calculation of pair corrections for 
Bhabha scattering has been done by LABSMC~\cite{arby}.

The conclusion for the inclusion of pair effects in the two-fermion
cross-section  are as follows:
with the exception of the $1.7$ per mill (tag of 
$1.1$ per mill) 
difference for inclusive hadrons, all theoretical uncertainties are 
well below the experimental precision tags.
Especially for the case of Bhabha scattering it would be 
highly desirable to have more than one code predicting the effects of 
secondary pairs. 
Note that improvements are still expected in GENTLE, TOPAZ0 and KKMC + KORALW.

In the following we will discuss items that are related to four-fermion
production.

\begin{itemize}
\item $\wb\wb$ signal: the CC03 class
\end{itemize}

While the CC03 cross-section is not an
observable, it is nevertheless a useful quantity at LEP~2 energies where it can
be classified as a pseudo-observable. It contains the interesting physics,
such as the non-abelian couplings and the sensitivity of the total 
cross-section to $\mw$ near the $\wb$-pair threshold.
The goal of this common definition is to be able to combine the
different final state measurements from different experiments so that the
new theoretical calculations can be checked with data at a level better than
$1\%$.

It is worth summarizing the status of the $\wb\wb$
cross-section prior to the 2000 Winter Conferences.
Nominally, any calculation for $\eeWWffff$
was a tree level calculation including as much as possible of the universal
corrections in some sort of Improved Born Approximation (IBA).
A CC03 cross-section, typically in the $\gf$-scheme,
with universal ISR QED and non-universal ISR/FSR QED corrections
produces a curve that been used for the definition of the standard model
prediction with a $\pm 2\%$ systematic error assigned to it.
However, we have clear indications that non-universal electroweak 
corrections for $\wb\wb$(CC03) cross-section are not small and even 
larger than the experimental LEP accuracy.
Furthermore, one should stress the importance of photon reconstruction
at LEP~2 accuracy.

Recently~\cite{4f}, a new electroweak $\ord{\alpha}$ CC03
cross-section has become
available, in the framework of double-pole approximation (DPA), showing a 
result that is $2.5 \div 3\%$ smaller than the old CC03 cross-section.
This is a big effect since the combined experimental accuracy of LEP 
experiments is even smaller.

\bfi
\centerline{\epsfig{file=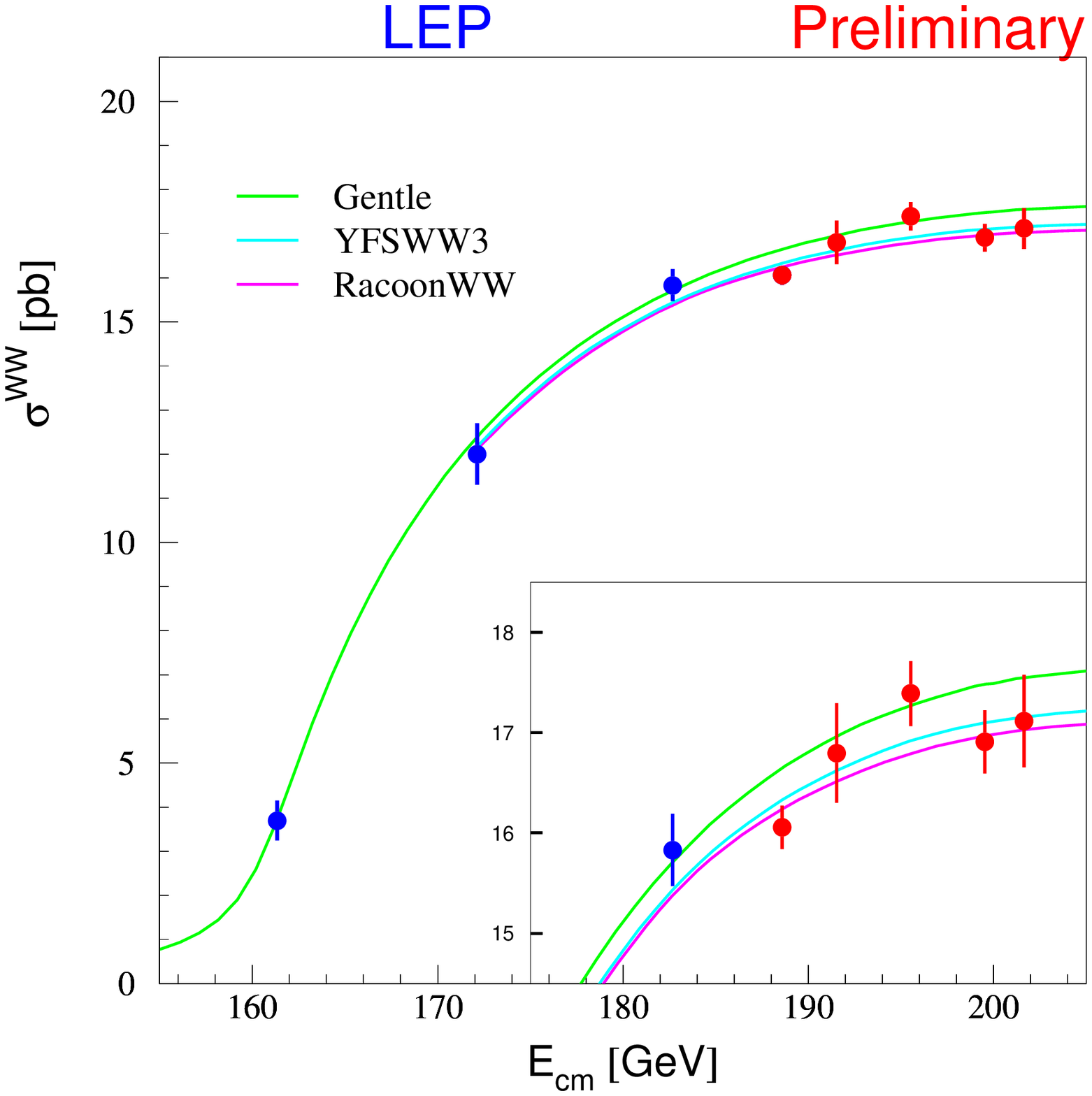,height=9cm,angle=0}}
\efi

DPA emerges from the CC03 diagrams upon projecting the $\wb$-boson momenta 
in the matrix element to their on-shell values. 
This means that the DPA is based on the residue of the 
double resonance, which is a gauge-invariant quantity. 
In contrast to the CC03 cross-section, the DPA
is theoretically well-defined. DPA provides a convenient framework
for the inclusion of radiative corrections, but should not be
applied for Born-level calculations.
Summarizing we may say that the at present only workable approach for 
evaluating the radiative corrections to resonance-pair-production processes, 
involves the so-called leading-pole approximation. This approximation 
restricts the complete pole-scheme expansion to the term with the
highest degree of resonance.

Conclusions for CC03 are as follows:
the data are in good agreement with the predictions of 
RacoonWW~\cite{racoonww} and YFSWW3~\cite{yfsww3} (see also BBC~\cite{bbc}).
At the time of Winter 2000 predictions of YFSWW3 were 
about $0.5\,\%-0.7\,\%$ higher, somewhat larger than intrinsic DPA
uncertainty.
The main source of this discrepancy is found, RacoonWW and
YFSWW3 differ only by about $0.3\,\%$ at LEP~2 energies for total
cross-sections and within $1\,\%$ in angular and invariant-mass distributions.
There is a general satisfaction with the progress induced by new
DPA calculations. Nevertheless, the theoretical uncertainty could  
probably be improved somewhat in the future.

\bfi
\centerline{\epsfig{file=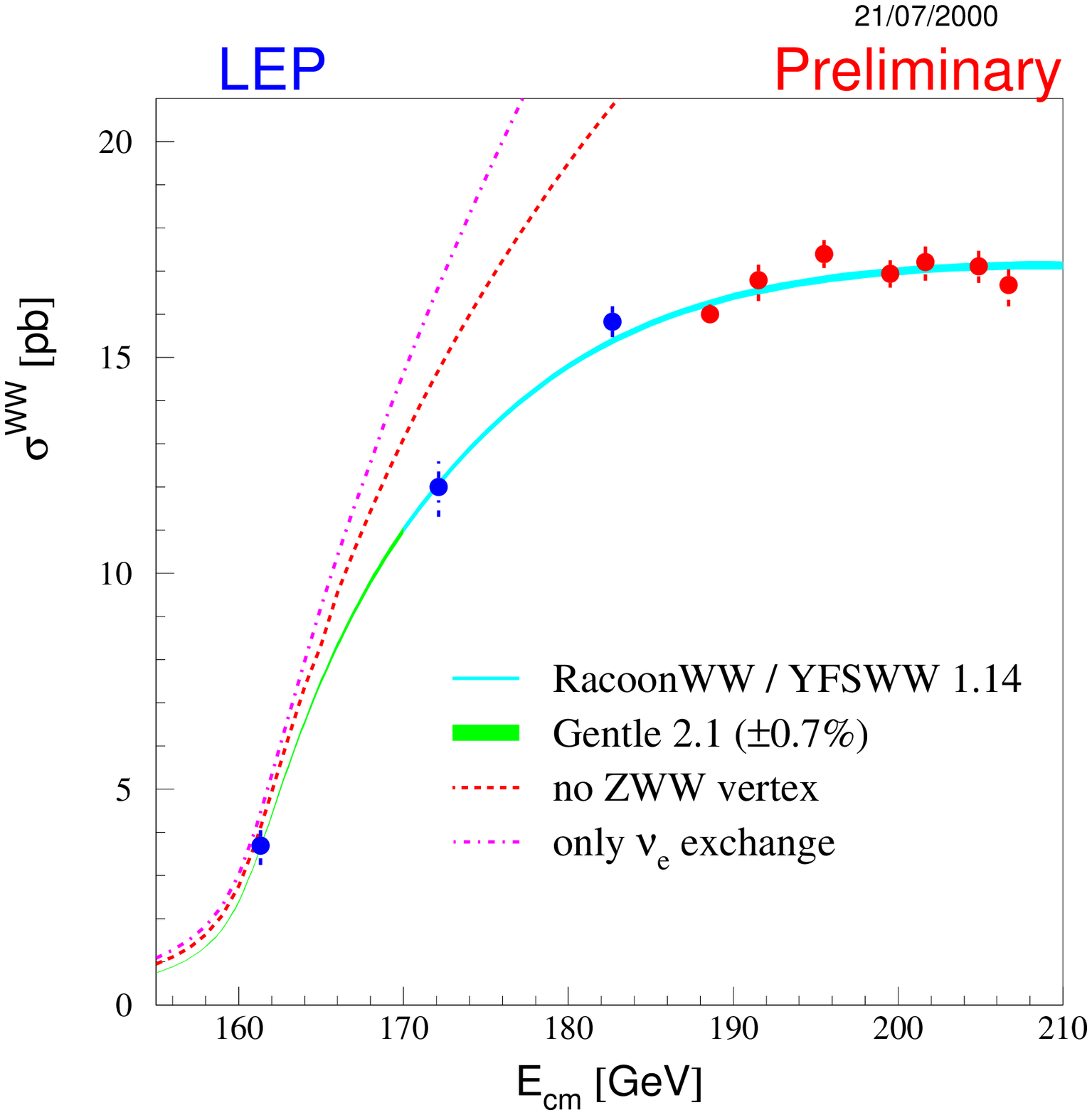,height=9cm,angle=0}}
\efi

\begin{itemize}
\item single-$\wb$ production
\end{itemize}

A fairly large amount of work has been done in the last
years on the topic of single-$\wb$ production.
The experimental community agreed on some setup to 
define the single-$\wb$ production and now this has been formalized 
in one of the {\tt LEP EWWG} meetings; there, it was decided to have a 
signal definition as follows:

\begin{enumerate}

\item $ee\nu\nu$, $t$-channel only, $E(e^+) > 20\,\GeV, \\
|\cos\theta(e^+)| < 0.95,  |\cos\theta(e-)| > 0.95$; 
\item $e\nu\mu\nu$,  $t$-channel only,  $E_(\mu^+)  > 20\,\GeV$; 
\item $e\nu\tau\nu$, $t$-channel only, $E_(\tau^+)  > 20\,\GeV$;
\item $e\nu u d$, $t$-channel only, $M(ud) > 45 GeV$;
\item $e\nu c s$, $t$-channel only, $M(cs) > 45 GeV$.
\end{enumerate}

The main problems in dealing with single-$\wb$ production are the 
correct choice of the energy scale in couplings and the proper
treatment of QED radiation in processes that are not dominated by 
annihilation diagrams..

For the energy scale in couplings we have now an exact calculation~\cite{fl} 
based on the massive formulation of the Fermion-Loop scheme (FL) which, at 
the Born-level (no QED)is known to be at the $1\,\%$ level of accuracy 
(see WTO~\cite{wto}). 
No program includes $\ord{\alpha}$ electroweak radiative corrections.
Note that the FL-scheme developed in~\cite{BHF1} and refined in
\cite{BHF2} makes the approximation of neglecting all masses for
the incoming and outgoing fermions in the processes $e^+e^- \to n\,$fermions.
The recent development, however, goes beyond this approximation.

A description of single-$\wb$ processes by means of the 
FL-scheme is mandatory because FL is the only known 
QFT consistent scheme that preserves gauge invariance and, moreover,
single-$\wb$ production is a process that depends on several
scales: the single-resonant $s$-channel exchange of $\wb$-bosons, 
the exchange of $\wb$-bosons in $t$-channel, the small scattering angle peak 
of outgoing electrons. 

A correct treatment of the multi-scale problem can only be achieved via 
FL-scheme and a naive rescaling cannot reproduce the full answer for
all situations, all kinematical cuts. 

The effect of QED on the total cross-sections are between $7\%$ and $10\%$ at 
LEP~2 energies. Furthermore, grc4f and SWAP~\cite{4f} have estimated that if 
one uses the wrong energy scale $s$ in the IS structure functions, the ISR 
effect is overestimated of about $4\,\%$.
SWAP~\cite{swap} estimates that the effects due to non-$s$-scales predict a 
lowering of the Born cross-section of about $8\,\%$.
SWAP results show a good agreement with those of grc4f~\cite{grace}. 

Conclusions for single-$\wb$ are as follows:
although we register substantial improvements upon the 
standard treatment of QED ISR, 
the problem is not yet fully solved for processes where the
non-annihilation component is relevant. A solution of it should rely 
on the complete calculation of the $\ord{\alpha}$ correction.
At the moment, a total upper bound of $\pm 5\,\%$ 
th. uncertainty should be assigned to single-$\wb$. 

We could say that QED in single-$\wb$ is 
understood at a level better than $4\,\%$ but we are presently unable 
to quantify this assertion. 

\begin{itemize}
\item $\zb\zb$ signal
\end{itemize}

NC02 is $e^+e^- \to \zb\zb$, ($t$ and $u$ channel), with all $\zb$ decay modes 
allowed. Since the interferences between the crossings are not double-resonant,
it is customary to consider them as background and to define the $\zb\zb$
signal from the absolute squares of the double-resonant diagrams only.
The choice is based on the observation that 
$R_{uucc/uuuu} = 2.06, R_{ddss/dddd} = 2.08$.

\bfi
\centerline{\epsfig{file=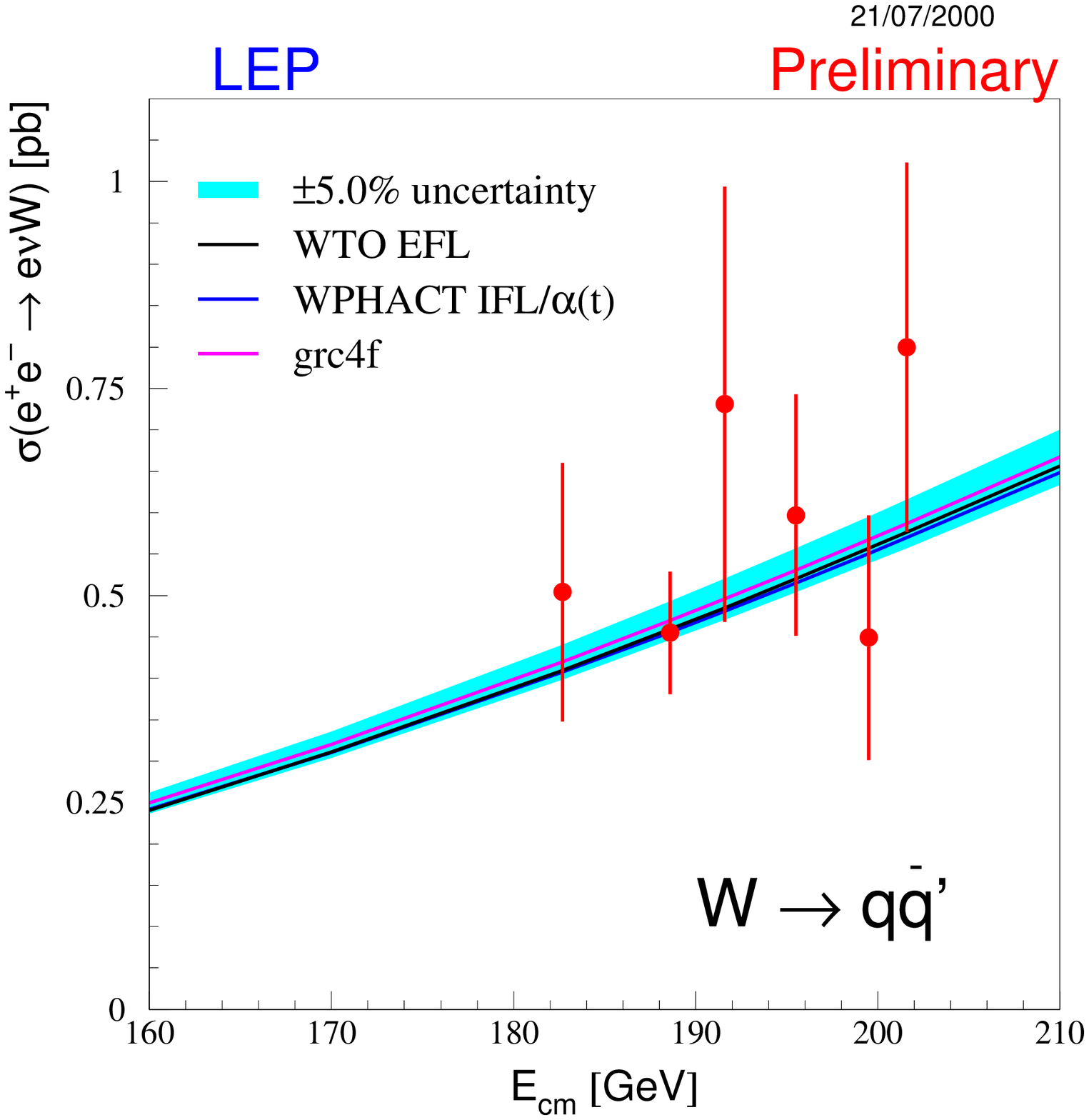,height=9cm,angle=0}}
\efi

Compared to the experimental uncertainty~\cite{sal} on the NC02 $\zb\zb$ 
cross-section
a difference of about $1\%$ between theoretical predictions is acceptable. 
The global estimate of theoretical uncertainty is $2\%$, again acceptable.
However, it would be nice to improve upon the existing calculations.

Conclusions for NC02 are as follows:
for the NC02 cross-section we have a $1\%$ 
variation, obtained by changing the input parameter set in GENTLE and in
ZZTO~\cite{zzto} and by varying from the standard GENTLE approach 
for ISR to the complete lowest order corrections. We estimate the real 
uncertainty to be $2\%$.
Furthermore, ZZTO which is a FL calculation (with universal ISR, 
FSR$^{}_{\rm QED}$ FSR${}_{\rm QCD}$ and running masses) agrees rather well 
with YFSZZ~\cite{yfszz}, roughly below the typical DPA accuracy of $0.5\%$, 
and the latter features leading pole approximation, on $\ord{\alpha^2}$ 
leading-logarithms YFS exponentiation (EEX).

The implementation of a DPA calculation, in more than 
one code, in the NC02 $\zb$-pair cross-section will bring the 
corresponding accuracy at the level of $0.5\%$, similar to the 
CC03 case. 

\bfi
\centerline{\epsfig{file=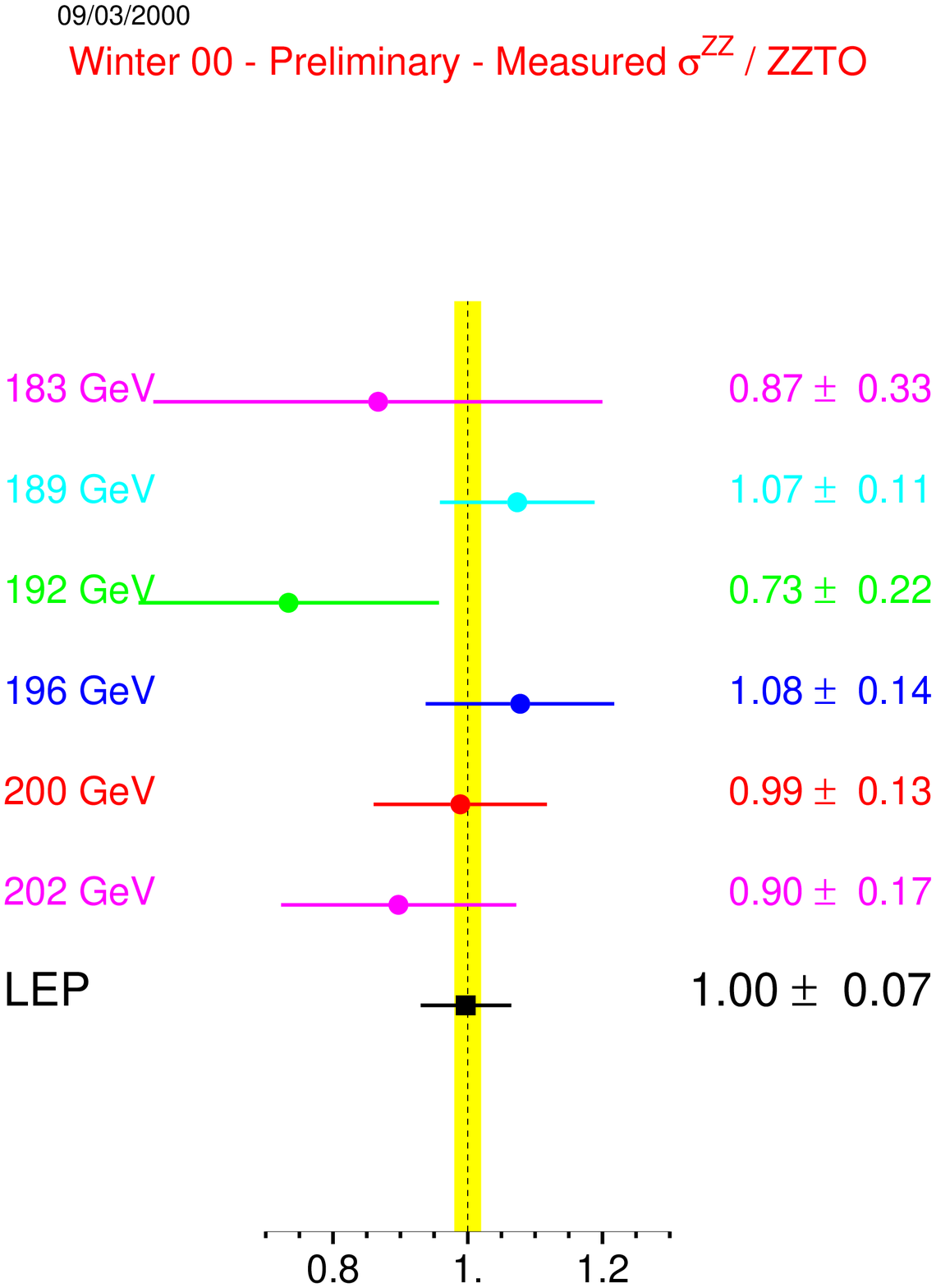,height=9cm,angle=0}}
\efi
 
\begin{itemize}
\item Conclusions
\end{itemize}

To gauge the priorities of this rather short summary one should remember that 
the experimental situation~\cite{gurtu} is rather different for $\wb\wb$ when 
compared to other processes. For
$\wb$-pairs, LEP (ADLO) is able to test the theory to below $1\%$, \ie, below
the old uncertainty of $\pm 2\%$ established in 1995.  Thus the
CC03-DPA, including non-leading electroweak corrections,
constitutes a very important theoretical development.
However, ADLO cannot test single-$\wb$ or $\zb\zb$-signal to an
equivalent level, since their total cross-section is of the order of
$1\,pb$ or less, $20$ times smaller than that of $\wb$-pair production.

\section*{Acknowledgments}
I would like to acknowledge the precious help and collaboration from all
participants in the LEP~2/MC Workshop and, in particular, from Martin 
Gr\"unewald. I thank W. Hollik for inviting me to this lively session
and the LEP EWWG for providing some of the figures.

\end{document}